\documentclass{emulateapj}

\usepackage{graphicx}
\usepackage{float} 
\usepackage{amsmath}
\usepackage{epsfig,floatflt}

\renewcommand{\d}[0]{\mathbf{d}}
\newcommand{\n}[0]{\mathbf{n}}

\renewcommand{\a}[0]{\mathbf{a}}

\newcommand{\B}[0]{\mathbf{B}}

\begin{document}

\title{The effect of systematics on polarized spectral indices}

\author{I. K. Wehus\altaffilmark{1}, U. Fuskeland\altaffilmark{2} and H. K. Eriksen\altaffilmark{2,3}}

\email{i.k.wehus@fys.uio.no}

\altaffiltext{1}{Astrophysics, University of Oxford, DWB, Keble Road, Oxford OX1 3RH, UK}

\altaffiltext{2}{Institute of Theoretical Astrophysics, University of
  Oslo, P.O.\ Box 1029 Blindern, N-0315 Oslo, Norway}

\altaffiltext{3}{Centre of Mathematics for Applications, University of
  Oslo, P.O.\ Box 1053 Blindern, N-0316 Oslo, Norway}


\begin{abstract}
We study four particularly bright polarized compact objects (Tau A,
Virgo A, 3C273 and Fornax A) in the 7-year WMAP sky maps, with the
goal of understanding potential systematics involved in estimation of
foreground spectral indices. We estimate the spectral index, the
polarization angle, the polarization fraction and apparent size and
shape of these objects when smoothed to a nominal resolution of
$1^{\circ}$ FWHM. Second, we compute the spectral index as a function
of polarization orientation, $\alpha$. Because these objects are
approximately point sources with constant polarization angle, this
function should be constant in the absence of systematics. However,
for the K- and Ka-band WMAP data we find strong index variations for
all four sources. For Tau A, we find a spectral index of
$\beta=-2.59\pm0.03$ for $\alpha=30^{\circ}$, and $\beta=-2.03\pm0.01$
for $\alpha=50^{\circ}$. On the other hand, the spectral index between
Ka- and Q-band is found to be stable. A simple elliptical Gaussian toy
model with parameters matching those observed in Tau A reproduces the
observed signal, and shows that the spectral index is in particular
sensitive to the detector polarization angle. Based on these findings,
we first conclude that estimation of spectral indices with the WMAP
K-band polarization data at $1^{\circ}$ scales is not robust. Second,
we note that these issues may be of concern for ground-based and
sub-orbital experiments that use the WMAP polarization measurements of
Tau A for calibration of gain and polarization angles.
\end{abstract}
\keywords{cosmic microwave background --- cosmology: observations --- methods: statistical}

\section{Introduction}
\label{sec:introduction}

One of the central goals in contemporary observational cosmology is to
detect the postulated background of primordial gravity waves predicted
by inflation. The most direct observational signature of these gravity
waves is a particular pattern in the polarization of the cosmic
microwave background (CMB) known as B-modes. The amplitude of these
gravity waves is typically parametrized in terms of the
tensor-to-scalar ratio, $r$ \citep[see, e.g.][ and references therein
  for a thorough review on inflation]{liddle:2000}. During the last
few years many experiments have been planned, built and fielded to
measure $r$, and the first relevant B-mode constraints have been
already published by BICEP ($r < 0.7$, Chiang et al. 2010) and QUIET
($r < 2.1$; QUIET 2011, 2012). Other ground-based and sub-orbital
experiments are expected to vastly improve on these limits in the very
near future.

In order to make an actual detection of the inflationary gravity
waves, it is widely believed that a sensitivity of $r\lesssim0.01$
will be required. In terms of map-domain sensitivity, this corresponds
to a signal with an RMS of a few tens of nK. Thus, not only will
exquisitely sensitive detectors be needed, but also detectors with
extremely low systematics.

However, the single most problematic systematic for future B-mode
experiments is likely not to come from the instrument itself, but
rather from the sky: Non-cosmological Galactic and extra-galactic
foregrounds, for instance synchrotron and thermal dust, radiates with
a temperature of several $\mu\textrm{K}$ on large angular scales in
the frequencies relevant for CMB measurements \citep[e.g.,][and
  references therein]{gold:2011}. Therefore, in order not
to be foreground dominated, these foregrounds must very likely be
suppressed by perhaps an order of magnitude or more. The only way to
achieve this is by making multifrequency observations of the same
fields of the sky, and exploit the different frequency dependency of
the various components to separate out the cosmological CMB signal
from the non-cosmological foregrounds.

As of today, a very large fraction of the information we have about
polarized foregrounds on large angular scales comes from the WMAP
satellite experiment, and in particular the lowest frequency channel
at 23 GHz (K-band). This map is routinely used both for studies of
foregrounds themselves, and as ancillary data for other
experiments. It is therefore critical to understand the
systematic limitations inherent in these data. In this paper we
measure the spectral indices of four particularly bright compact
objects (Tau A, Virgo A, 3C273 and Fornax A), with the goal of
understanding some of the issued involved in spectral index estimation
for CMB data in general: By considering high signal-to-noise objects
with known properties, we have a clear a-priori prediction, and
deviations from these expectations would indicate either model
problems or systematic errors.

\section{Data and model}
\label{sec:data}

\paragraph{Sky maps and processing} In this paper we consider the 7-year WMAP sky maps
\citep{jarosik:2011}, coadded over years and pixelized at a
HEALPix\footnote{http://healpix.jpl.nasa.gov} resolution of
$N_{\textrm{side}}=512$, corresponding to $7'$ pixels. These data are
available from LAMBDA\footnote{http://lambda.gsfc.nasa.gov}, including
all necessary ancillary data, such as beam profiles and noise
model. Most of our analysis is performed with the K- and Ka-band data,
although in one particular case we also consider the Q-band data. All
analyses are carried out in antenna temperature units, and given
that we will consider objects with steep synchrotron-like spectra we
adopt effective frequencies of 22.45 GHz (K-band), 32.64 GHz (Ka-band)
and 40.50 GHz (Q-band), respectively \citep{page:2003}.

Before one can estimate spectral indices across frequencies, it is
necessary to bring all maps to a common angular resolution. We
therefore smooth all maps to an effective resolution of $1^{\circ}$
FWHM by first deconvolving the instrument beam and then convolving
with a Gaussian beam of the desired size. Note that the smoothing
scale of $1^{\circ}$ is a particularly common value adopted in the
literature, and the results presented here are therefore of wide
interest.

Estimation of uncertainties for all scalar quantities is done by
forward Monte Carlo simulations. That is, we add smoothed noise
realizations to the actual WMAP data based on the provided noise
model, evaluate each statistic for each simulation, and then compute
the resulting standard deviation over the ensemble. Although there
already is a noise component present in the WMAP data, this is
identical for all simulations, and therefore do not contribute to the
variance. We emphasize, though, that uncertainties estimated in this
manner are only statistical in nature, and do not account for
systematic errors. 

\paragraph{Data selection} In this paper we consider the four
particularly bright point sources listed in Table
\ref{tab:image}. These were selected by thresholding the K-band
polarization map, $P = \sqrt{Q^2+U^2}$, at $100\mu\textrm{K}$, and
discarding all regions that either show obviously extended
features or have a strong background. This left us with Tau A as the
only near-Galactic source, and three high-latitude sources (Virgo A,
3C273 and Fornax A). For further details on the polarization
properties of these objects, see, e.g., \citet{aumont:2010,
  weiland:2011, fomalont:1989, ekers:1983, rottmann:1996}.

Only pixels in a $1^{\circ}$ radius disc around each source were kept
for analysis, although we also tried $2^{\circ}$ discs, obtaining
consistent, but slightly more noisy, results. Note that a Gaussian
beam of $1^{\circ}$ FWHM ($\sim2.35\sigma$) has dropped off to 6\% of
its peak value at a distance of $1^{\circ}$, and most of the volume is
therefore contained within this radius.

\paragraph{Data model}
The low-frequency WMAP polarization observations are strongly
dominated by synchrotron emission which has
a sharply falling spectrum. We therefore
approximate the total sky signal by a single power-law, resulting in
the following data model,
\begin{equation}
\d_{\nu} = \B \a \left(\frac{\nu}{\nu_0}\right)^{\beta} + \n_{\nu}.
\end{equation}
Here $\d_{\nu}$ is an $N_{\textrm{pix}}\times 3$ matrix listing the
Stokes' $I$, $Q$ and $U$ parameters for all relevant pixels at
frequency $\nu$ columnwise, $\B$ is a
$3N_{\textrm{pix}}\times3N_{\textrm{pix}}$ matrix denoting convolution
with the common instrumental beam, $\a$ denotes the true sky signal
amplitude as measured at a reference frequency $\nu_0$, and $\n_{\nu}$
is (smoothed) instrumental noise. All values are defined in antenna
temperature units. Coordinates are defined according to the HEALPix
convention \citep{gorski:2005}.

\begin{figure}[t]
\begin{center}
\mbox{\epsfig{figure=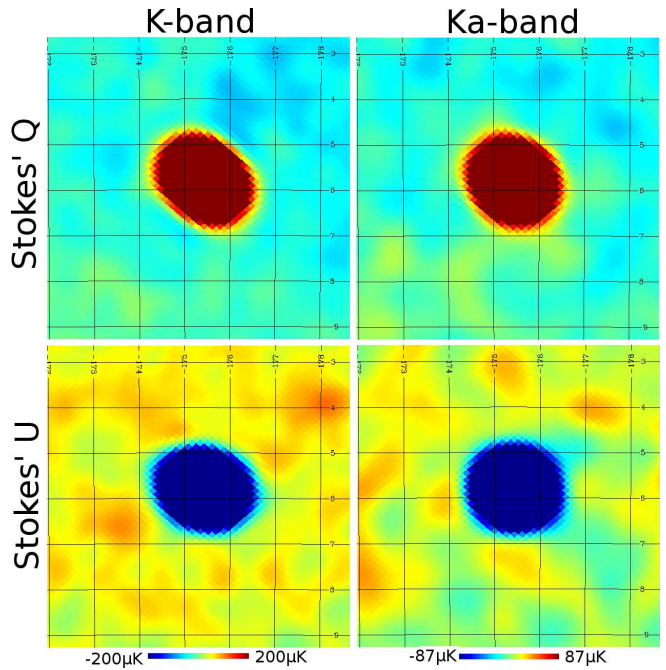,width=\linewidth,clip=}}
\end{center}
\caption{Tau A as observed by the two lowest WMAP frequencies, and
  rotated into a coordinate system offset by $22.5^{\circ}$ from the
  intrinsic polarization direction of Tau A. Note both that the K-band
  images are significantly more asymmetric than the Ka-band images,
  and also that the two Stokes' parameters within a single band are
  significantly different.}
\label{fig:objects}
\end{figure}

\section{Methods}
\label{sec:methods}

Given the data and model described in Section \ref{sec:data}, we
estimate the polarization angle and fraction, the spectral index and
the apparent shape of each source. First, we note that each of the
four objects considered here are well known in the literature, and have known
polarization properties. Further, they are all known to be much
smaller than $1^{\circ}$ in angular dimensions (see Table
\ref{tab:image} for precise details), and their polarization angles
are known to be quite stable as a function of frequency. (For example,
the polarization angle of Tau A is known to vary by only a few degrees
over more than 10 decades in frequency; e.g., Aumont et al. 2010.) We
therefore assume that there is no real substructure within each source
on the scales we consider.

\begin{deluxetable*}{lccccccccc}
\tablewidth{0pt}
\tablecaption{\label{tab:image}Apparent object position, size and shape}
\tablecomments{These beam parameters are derived in the
  coordinate system defined by the polarization angle of the
  respective source. Only statistical errors are included in the
  uncertainties, not systematic errors.}
\tablecolumns{10}
\tablehead{  & \emph{Longitude}  & \emph{Latitude} & \emph{Size}  & \multicolumn{2}{c}{\emph{FWHM} (degrees)} & \multicolumn{2}{c}{\emph{Ellipticity}} &  \multicolumn{2}{c}{\emph{Orientation} (degrees)}   \\
\emph{Object}       & (degrees) & (degrees) & (arcmin)    &  K & Ka                 & K & Ka & K & Ka}
\startdata
Tau A         & $184.56$ & $\,-5.78$ & $7 \times 5$              & $0.985\pm0.001$ & $0.992\pm0.001$& $0.142\pm0.001$ & $0.079\pm0.001$& $54.0\pm0.3$& $57.8\pm1.5$\\
Virgo A & $283.78$ & $\phm\phd\phd74.49$ & $8 \times 6$          & $0.91\pm0.02$ & $0.89\pm0.05$& $0.17\pm0.04$ & $0.13\pm0.06$& $76\pm9$& $56\pm33$\\
3C273                & $289.95$ & $\phantom{1}64.36$ & $< 1$  & $1.05\pm0.02$ & $0.88\pm0.05$& $0.14\pm0.03$ & $0.22\pm0.07$& $117\pm7$& $97\pm15$\\
Fornax A             & $240.16$ & $-56.69$ & $12 \times 9$       & $1.09\pm0.02$ & $1.09\pm0.04$& $0.24\pm0.02$ & $0.24\pm0.07$& $9\pm3$& $5\pm60$
\enddata
\end{deluxetable*}

\paragraph{Polarization angle and fraction} Since our objects
effectively are point sources with constant polarization angle, there
should (ideally) be a single well-defined coordinate system in which
all signal is aligned with the Stokes' $Q$ parameter. We search for
this direction, $\alpha$, by minimizing the signal in the
corresponding $U$ parameter,
\begin{equation}
\chi^2(\alpha) = \sum_{p} \left(\frac{-Q_p \sin 2\alpha + U_p \cos 2\alpha}{\sigma}\right)^2,
\end{equation}
where $Q_p$ and $U_p$ are the Stokes' parameters in Galactic
coordinates, and $\sigma$ is the noise level. Having rotated the data
into the intrinsic polarization direction of the source, the
polarization fraction is found simply by $\Pi =
\left<Q(\alpha)/T\right>$.

\begin{figure}[t]
\begin{center}
\mbox{\epsfig{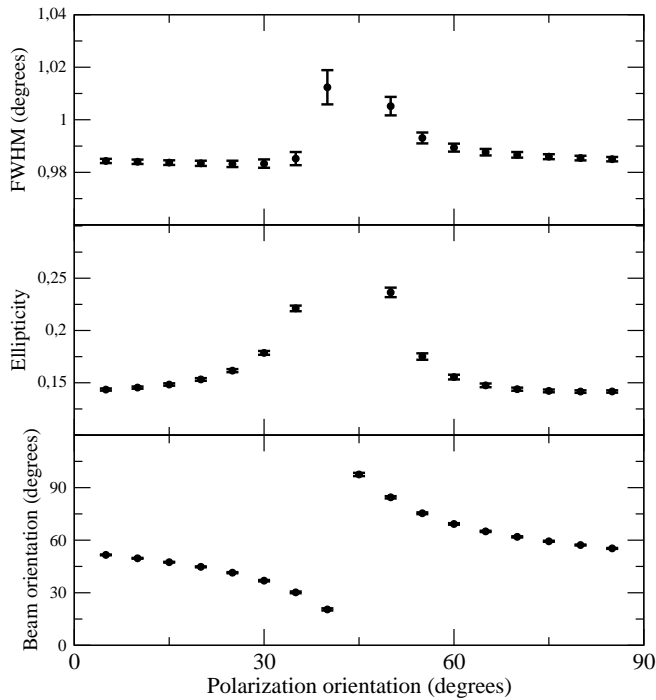}}
\end{center}
\caption{Beam parameters computed from Tau A at K-band as a function
  of polarization orientation. Note that all uncertainties include
  statistical errors only, not systematic.}
\label{fig:beam_vs_psi}
\end{figure}

\begin{figure}[t]
\begin{center}
\mbox{\epsfig{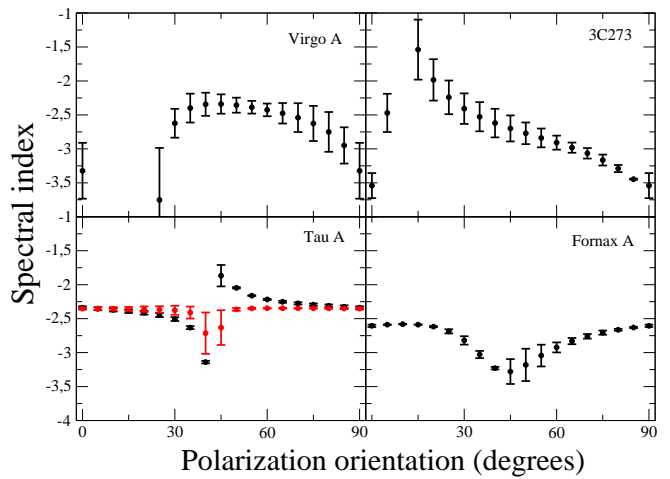}}
\end{center}
\caption{Spectral index as a function of polarization angle for the
  four objects considered in this paper. The black points are computed
  from the K- and Ka-bands; the red points (for Tau A only) are
  computed from the Ka- and Q-bands.}
\label{fig:beta_vs_psi}
\end{figure}

\begin{figure}[t]
\begin{center}
\mbox{\epsfig{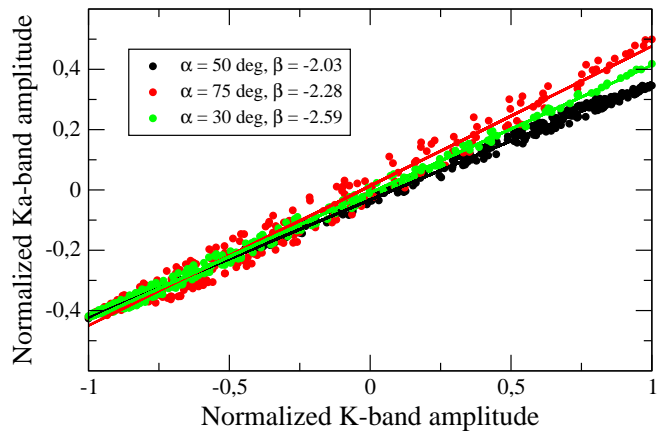}}
\end{center}
\caption{TT plots for Tau A for three different polarization
  orientations. Note that the slopes are different for each direction,
  corresponding directly to different effective spectral
  indices. Further, there is no sign of either instrumental noise or
  background, demonstrating that the results are highly robust against
  such effects.}
\label{fig:indfit}
\end{figure}

\paragraph{Observed ellipticity and FWHM} Although we
smooth the data to a common angular resolution, and therefore should
expect that the observations to have the desired FWHM, this is
not true in practice due to beam asymmetries. To study the effective
beam as a function of Stokes' parameters, we rotate the original map
by a rotation angle $\alpha$ into a new coordinate system
$Q(\alpha) = Q\cos(2\alpha)+U\sin(2\alpha)$, and consider all angles
between 0 and $90^{\circ}$ in steps of $5^{\circ}$. Then, in this new
coordinate system we fit an elliptical Gaussian, $g(Q_0,
\textrm{FWHM}, \epsilon, \psi)$, to the $Q$ signal by minimizing
\begin{equation}
\chi^2 = \sum_{p} \left(\frac{Q_p(\alpha) -
  g(Q_0, \textrm{FWHM}, \epsilon, \psi)}{\sigma}\right)^2,
\end{equation}
where $Q_0$ is the source amplitude, $\epsilon$ is the
ellipticity, and $\psi$ is the direction of the major semiaxis. (Note
that it is sufficient to consider only the $Q$ component, because we
rotate through all angles $\alpha$. Thus, $\alpha=45^{\circ}$
corresponds to $U$ in the original system.)

\begin{figure}[t]
\begin{center}
\mbox{\epsfig{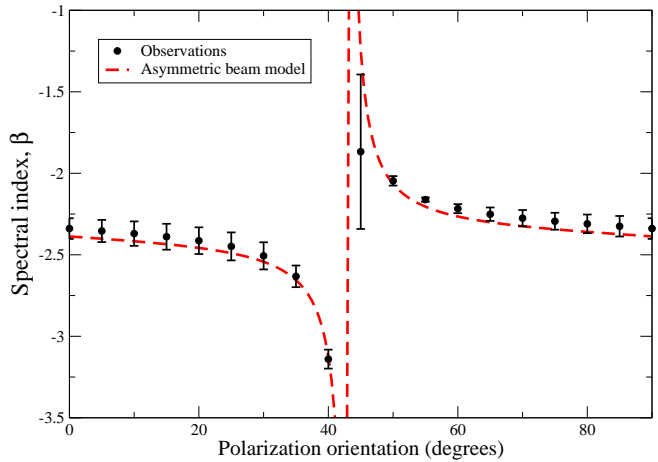}}
\end{center}
\caption{Comparison of the observed (black points) and simulated (red
  line) spectral index for Tau A. The simulation is based on a
  noiseless point source observed with the same beam and polarization
  parameters as measured for Tau A. (See Tables \ref{tab:image} and
  \ref{tab:polarization} for full specification.)}
\label{fig:beta_vs_psi_sim}
\end{figure}

\paragraph{Spectral indices} Finally, we estimate spectral indices for
both $Q(\alpha)$ and $U(\alpha)$ using a standard TT-plot
approach. For a single pixel with noiseless data, this approach is
simply defined by
\begin{equation}
\label{eq:tt_ideal}
\frac{d_{\nu_1}(p)}{d_{\nu_2}(p)} =
\left(\frac{\nu_1}{\nu_2}\right)^{\beta} \Rightarrow \beta =
\frac{\log [d_{\nu_1}(p)/d_{\nu_2}(p)]}{\log [\nu_1/\nu_2]}.
\end{equation}
However, for multiple noisy observations more robust results are
obtained by fitting a straight line, $y = ax+b$, to $d_{\nu_1}$ as a
function of $d_{\nu_2}$. An additional advantage of this method is its
insensitivity to absolute offsets in the data. The spectral index is
given as $\beta = \log a/\log(\nu_1/\nu_2)$. Since both $d_{\nu_1}$
and $d_{\nu_2}$ have uncertainties, it is important to use a method
that supports uncertainties in both directions with making the linear
fit. In this paper, we adopt the approach described by \citet{petrolini:2011}.
As in the case of the beam parameters, we also compute the spectral
index as a function of polarization angle. 

\begin{deluxetable*}{lcccccc}
\tablewidth{0pt}
\tablecaption{\label{tab:polarization}Polarization properties }
\tablecomments{Uncertainties on polarization fraction and angles
  include only statistical errors; uncertainties on spectral indices
  additionally include an estimate of systematic errors.}
\tablecolumns{7}
\tablehead{  & \multicolumn{2}{c}{\emph{Polarization fraction}}  & \multicolumn{2}{c}{\emph{Polarization angle} (degrees)} &   \multicolumn{2}{c}{\emph{Spectral index}} \\
Object &   K & Ka & K & Ka & $\beta_T$ & $\beta_{P}$}
\startdata
Tau A          & $6.17\pm0.01$ & $6.48\pm0.04$& $88.43\pm0.03$& $87.6\pm0.1$& $-2.280\pm0.001$ & $-2.33\pm0.01$ \\
Virgo A        & $3.4\pm0.1$  & $5.1\pm0.8$ & $-27\pm1$        & $-24\pm4$      & $-2.62\pm0.01$   & $-2.5\pm0.3$ \\
3C273          & $5.8\pm0.2$  & $4.4\pm0.6$ & $52.7\pm0.6$  & $44\pm2$    & $-2.27\pm0.01$ & $-2.8\pm0.2$ \\
Fornax A       & $6.7\pm0.2$ & $6.7\pm0.5$& $-2.6\pm0.7$& $-5\pm2$& $-2.90\pm0.02$ & $-2.6\pm0.2$ 
\enddata
\end{deluxetable*}

Estimation of uncertainties for spectral indices is a non-trivial
issue, because the inherent systematic errors turn out to be
significantly larger than the statistical. We therefore add a
systematic error term in quadrature to the statistical error. The
systematic error is defined by splitting the observed data points in
two sets, according to whether $d_{\nu_1}$ is larger or smaller than
$a d_{\nu_2} + b$, and estimate a new slope for each set. The
systematic error is taken to be the half difference between the two
slopes.

\section{Results}
\label{sec:wmap}

In Table \ref{tab:image} we list the position and apparent
(beam-convolved) size and shape of each of the four objects under
consideration. The polarization fraction and angles, and spectral
indices are tabulated in Table \ref{tab:polarization}. Images of Tau A
are shown in Figure \ref{fig:objects}, both for K- (left column) and
Ka-band (right column), and for Stokes' $Q$ and $U$ parameters. In
order to highlight the beam differences between these cases, we have
first adopted a coordinate system which is offset by $22.5^{\circ}$
from the intrinsic polarization direction of Tau A. This ensures a
significant signal-to-noise in both $Q$ and $U$. Second, the color
scale is tuned to highlight the tails of the instrumental beam, and
scaled properly between the two frequencies taking into
account the spectral index of Tau A.

The main results of this paper are shown in Figures
\ref{fig:beam_vs_psi} and \ref{fig:beta_vs_psi}. The first figure
shows the beam parameters for Tau A as a function of polarization
orientation, and the second shows the spectral index as a function of
polarization direction for all four sources. In the latter, the black
points indicate the spectral index computed from K- and Ka-bands, and
(for Tau A only) the red points show the spectral index between Ka-
and Q-band. Figure \ref{fig:indfit} shows a subset of the Tau A TT
plots that are used for the K-Ka calculations, corresponding to
$\alpha=30^{\circ}$, $50^{\circ}$ and $75^{\circ}$, respectively.

As seen from the results shown in Figure \ref{fig:beta_vs_psi}, the
polarized spectral index as measured by WMAP between K- and Ka-band at
$1^{\circ}$ angular scale depends strongly on the coordinate system in
which the index is computed. For Tau A, the derived index varies
between, say, $\beta = -2.6$ for a rotation angle of
$\alpha=30^{\circ}$ and $\beta=-2.0$ for $\alpha=50^{\circ}$.  This
effect is statistically highly significant, and it is robust with
respect to instrumental noise and background levels (see Figure
\ref{fig:indfit}). It therefore indicates the presence of a real
systematic effect not taken into account in the present analysis.

To understand these structures in greater detail, we construct an
elliptical Gaussian model of Tau A based on the parameters listed in
Tables \ref{tab:image} and \ref{tab:polarization} at K- and Ka-band,
and estimate the spectral index from the resulting noiseless model, as
for the real data. The results are shown in Figure
\ref{fig:beta_vs_psi_sim}. Clearly, the model faithfully reproduces
the observed structures. The only difference is a slight vertical
offset, which is due to the fact that the measured spectral index
(reported in Table \ref{tab:polarization}) is not a perfectly unbiased
estimate of the true spectral index in the presence of systematics.

\begin{figure}
\begin{center}
\mbox{\epsfig{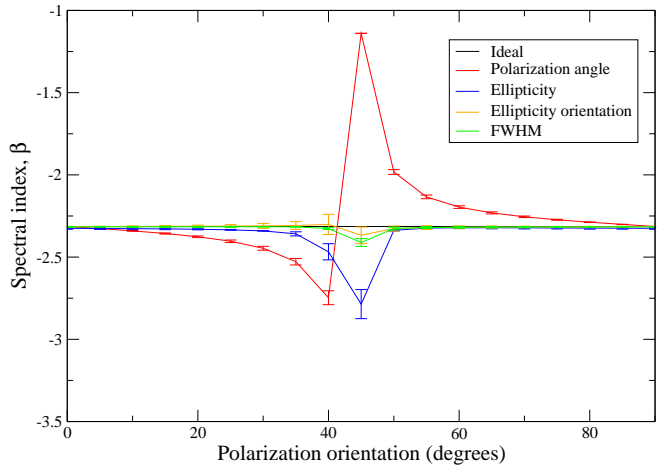}}
\end{center}
\caption{The effect of various systematics on the measured spectral
  index of Tau A between K and Ka-band. The simulated K-band map is
  always based on the measured Tau A parameters. For the black curve
  we have used the same beam and polarization angle parameters for Ka
  as for K, corresponding to an ideal instrument. For the other curves
  we change one parameter at a time to the measured Ka value.}
\label{fig:beta_vs_psi_systematics}
\end{figure}

We can now use this model to understand the relative importance of the
various systematic effects. To do so, we start out with an ideal
model, adopting the observed K-band parameters also for Ka-band, and
set the Ka parameters one-by-one to their true values. The results
from this exercise are shown in Figure
\ref{fig:beta_vs_psi_systematics}. Here we see that the most important
systematic by far is the detector angle, and this effect alone
reproduces the signal seen in Figure \ref{fig:beta_vs_psi_sim} very
well. The second most important effect is the beam ellipticity, which
is at least three to four times smaller than the detector angle effect
over most of the well-sampled regions of the polarization
orientation. Other effects are small compared to these two.

\section{Conclusions}
\label{sec:discussion}

We have studied four particularly bright polarized point sources in
the 7-year WMAP data, with the goal of understanding the effect of
systematics on polarized spectral index estimation. This topic is
important for at least two reasons. First, the WMAP polarization sky
maps represents the best currently available full-sky measurements of
the polarized foregrounds at CMB frequencies. As a result, they play a
critical role in the analysis and optimization of existing and future
B-mode experiments.  Second, many ground-based and sub-orbital
experiments use the WMAP polarization measurements of Tau A directly
as a calibration source for both detector angles and absolute gain. 

In this paper, we have found that the observed polarized spectral
index of the relevant sources depends sensitively on the coordinate
system in which the index is estimated. For example, the spectral
index of Tau A is $\beta=-2.59\pm0.03$ for a coordinate system rotated
by $30^{\circ}$ relative to the intrinsic polarization direction of
the source, while it is $\beta=-2.03\pm0.01$ in a coordinate system
rotated by $50^{\circ}$. The most significant contributor to this
effect is the slightly different polarization angles of the K- and
Ka-band detectors, with some smaller contribution coming from beam
asymmetries. Experiments that, directly or indirectly, use the K- and
Ka-band measurements of Tau A as a calibrator source should take into
account these systematic uncertainties when performing their analyses.

Finally, we note that the test described in this paper is very simple
to implement, only takes a few CPU seconds to run, and have a very
direct and intuitive interpretation. We therefore expect other
experiments to find it useful as a test of their own systematics, in
particular when applied to Tau A.

\begin{acknowledgements}
This project was supported by the ERC Starting Grant StG2010-257080
and a Leverhulme visiting professorship for HKE. Some of the results
in this paper have been derived using the HEALPix \citep{gorski:2005}
software and analysis package.
\end{acknowledgements}

\end{document}